\documentclass[%
 preprint,
groupedaddress,
superscriptaddress,
 amsmath,amssymb,
 aps,
 prl,
 showpacs
]{revtex4-2}
\usepackage{graphicx}
\usepackage{makecell,booktabs}
\usepackage{dcolumn}
\usepackage{bm}
\usepackage{hyperref}
\hypersetup{
    colorlinks=true,
    urlcolor= blue,
    citecolor=blue,
linkcolor= blue}
\usepackage{xcolor}

\begin{document}

\title{Cryogenic growth of aluminum: structural morphology, optical properties, superconductivity and microwave dielectric loss}

\author{Wilson J. Y\'{a}nez-Parre\~{n}o}
\affiliation{Department of Electrical and Computer Engineering, University of California, Santa Barbara, California 93106, USA\looseness=-1}
\author{Teun A. J. van Schijndel}
\affiliation{Department of Electrical and Computer Engineering, University of California, Santa Barbara, California 93106, USA\looseness=-1}
\author{Anthony P. McFadden}
\affiliation{National Institute of Standards and Technology, Boulder, Colorado 80305,  USA\looseness=-1}
\author{Kaixuan Ji}

\affiliation{National Institute of Standards and Technology, Boulder, Colorado 80305,  USA\looseness=-1}
\affiliation{Department of Physics, University of Colorado, Boulder, Colorado 80309, USA\looseness=-1}

\author{Susheng Tan}
\affiliation{Department of Electrical and Computer Engineering, University of Pittsburgh, Pittsburgh, PA 15260, USA\looseness=-1}
\affiliation{Petersen Institute of Nanoscience and Engineering, University of Pittsburgh, Pittsburgh, PA 15260, USA\looseness=-1}
\author{Yu Wu}
\affiliation{Department of Electrical and Computer Engineering, University of California, Santa Barbara, California 93106, USA\looseness=-1}
\author{Sergey Frolov}
\affiliation{Department of Physics and Astronomy, University of Pittsburgh, Pittsburgh, Pennsylvania 15260, USA\looseness=-1}

\author{Stefan Zollner}
\affiliation{Department of Physics, New Mexico State University, Las Cruces, New Mexico 88003, USA\looseness=-1}

\author{Raymond W. Simmonds}
\affiliation{National Institute of Standards and Technology, Boulder, Colorado 80305, USA\looseness=-1}

\author{Christopher J. Palmstr\o{}m}%
\email[]{cjpalm@ucsb.edu}
\affiliation{Department of Electrical and Computer Engineering, University of California, Santa Barbara, California 93106, USA\looseness=-1}
\affiliation{Materials Department, University of California, Santa Barbara, California 93106, USA\looseness=-1}

\date{\today}

\begin{abstract}

We explore the molecular beam epitaxy synthesis of superconducting aluminum thin films grown on c-plane sapphire substrates at cryogenic temperatures of 6 K and compare their behavior with films synthesized at room temperature. We demonstrate that cryogenic growth increases structural disorder, producing crystalline grains that modify the optical, electrical, and superconducting properties of aluminum. We observe that cryogenic deposition changes the color of aluminum from fully reflective to yellow and correlate the pseudo-dielectric function and reflectance with structural changes in the film. We find that smaller grain sizes enhance the superconductivity of aluminum, increasing its critical temperature and critical field. We then estimate the superconducting gap and coherence length of Cooper pairs in aluminum in the presence of disorder. Finally, we fabricate superconducting microwave resonators on these films and find that, independently of the growth temperature, the system is dominated by two-level system loss with similar quality factors in the high and low power regimes. We further measure a higher kinetic inductance in the cryogenically grown films.

 \end{abstract}
\maketitle

\section{\label{sec:level1}Introduction}
Disorder in a superconductor can dramatically change its physical properties \cite{ANDERSON195926,PhysRevB.3.1611,PhysRevB.98.184510, PhysRev.115.1597,PhysRev.168.444}. Some of these changes might be favorable for technology, such as the ability to tune the critical temperature and critical field with disorder \cite{LYNTON1957165, PhysRev.168.444, PhysRev.115.1597, PhysRevApplied.11.011003, 10.1063/5.0017749, Gao2025, 10.1063/5.0250146Dome} and the presence of significant kinetic inductance and Kerr nonlinearity that can be advantageous for superconducting nonlinear devices like single-photon detectors, parametric amplifiers and quantum bits (qubits) \cite{10.1063/1.4737408, 10.1063/5.0017749, PhysRevLett.122.010504,10.1063/5.0100961,Gao2025, PhysRevLett.121.117001, Grunhaupt2019, PhysRevLett.133.260604,PhysRevApplied.20.044021}. However, it is not clear to what extent disorder and grain size in a superconductor affects the performance of quantum information systems and quantum circuits. Recently, a report has suggested universal scaling between microwave dissipation and superfluid density as an intrinsic bulk dissipation channel in superconductors \cite{charpentier2025}. Moreover, microwave dielectric loss at surfaces and interfaces has been attributed as one of the main sources of decoherence in these systems \cite{10.1063/5.0017378McRae,doi:10.1126/science.abb2823}, and there is a significant ongoing effort to characterize materials quality as a way to understand and minimize decoherence effects in microwave circuits in the context of two-level system loss \cite{10.1063/5.0017378McRae, 4757203McDermott, 10.1063/1.3693409,doi:10.1126/science.abb2823,Place2021,PhysRevX.13.041005, Wang2022,bland20252dtransmons1ms, PhysRevApplied.23.034025,McFadden2025,mcfadden2025NbSi}. For this reason, in this work we focus on the synthesis of aluminum films grown on c-plane sapphire substrates under ultra high-vacuum conditions and use a cryogenic environment during deposition as a way to introduce structural disorder in the form of grain size without the presence of unintentional doping that can affect its physical properties in an uncontrolled manner. We then study structural morphology, optical properties, superconductivity and microwave dielectric loss down to the single-photon regime.

Aluminum has several favorable qualities that makes it the material of choice for superconducting quantum information applications. From a practical perspective, it has the advantage of growing epitaxially on sapphire ($\mathrm{Al_2O_3}$) \cite{10.1063/5.0017378McRae,10.1063/1.3693409,Richardson2016-jd}, silicon (Si) \cite{10.1063/5.0017378McRae,mohseni2025,10.1116/1.4971200,Richardson2016-jd} and III-V substrates (InAs, GaAs, $\mathrm{In_{x}Ga_{1-x}As}$ and others) \cite{10.1063/1.328722,10.1063/5.0029855,10.1063/5.0023743,PhysRevB.93.155402,10.1116/1.5145073,Krogstrup2015}. It has a natural self-limiting oxide ($\mathrm{AlO_x}$) that expands during oxidation which allows the fabrication of pin-hole free Josephson junctions \cite{PhysRevMaterials.8.046202,10.1063/1.2977725}. It is widely used in the semiconductor industry with mature nanofabrication protocols and well established etching chemistries \cite{mohseni2025}. In addition, it has been shown that aluminum deposition at substrate temperatures below 293 K and down to liquid nitrogen temperatures improves the aluminum film quality and can lead to high transparency in superconductor/semiconductor heterostructures \cite{Krogstrup2015,PhysRevB.93.155402}. Thus it is interesting to investigate deposition at even lower temperatures. From a fundamental perspective, aluminum is interesting because it has a $\mathrm{181\ \mu eV}$ superconducting gap that increases with decreasing thickness \cite{doi:10.1126/sciadv.adf5500}. It has a large Fermi velocity of $\mathrm{2\times10^6\ m/s}$ which results in a large coherence length of Cooper pairs in the clean limit that can be as high as $\mathrm{1.6\ \mu m}$ \cite{ashcroft_solid_1976,PhysRev.168.444,Hart_1997}. Superconductivity in aluminum is extremely robust against disorder and has shown to survive even in the granular regime in which aluminum domains in an oxide matrix show superconductivity with a critical dome that can be tuned by oxidation and domain size due to confinement \cite{PhysRev.168.444,Grunhaupt2019, 10.1063/5.0250146Dome, PhysRevLett.100.187001,Bose2014ua,PhysRevB.93.100503}. Finally, like most metals, aluminum shows crystalline domains even when grown on amorphous substrates (supplemental material). This gives the opportunity to study growth kinetics and crystal formation in the presence of disorder. 



Aluminum was grown on sapphire ($\mathrm{Al_2O_3}\ (0001)$) substrates because they have low dielectric loss with a bulk loss tangent that can be as low as $tan(\delta_i)=19(6)\times10^{-9}$ \cite{PhysRevApplied.19.034064}. In the c-plane of sapphire (Fig. 1(a)), the distance between aluminum and oxygen atoms ($\mathrm{2.38\ \AA}$) is closely matched with the interatomic distance of aluminum atoms in the (111) plane of the aluminum film ($\mathrm{2.34\ \AA}$). This corresponds to a lattice constant mismatch of 1.6$\%$ in which the aluminum film grows under tensile strain. We expect this value to be small enough to produce epitaxy when the film is grown at room temperature and low enough to introduce disorder without completely destroying crystallinity when the film is grown at cryogenic temperature. We estimate the critical thickness of aluminum to be 6 nm, which means that in thicker films we expect to have dislocations \cite{MATTHEWS1974118,10.1063/1.97637}. The precise kinetics of growth at different temperatures can be complex, but we expect cryogenic growth to freeze dislocation motion in the film. In general, we also expect cryogenic growth to minimize the energy of the adatoms in the surface of the sample during deposition which will increase the sticking probability, minimize surface diffusion, and produce chemically abrupt and pristine interfaces.

\section{\label{sec:level12}Results}
\subsection{\label{sec:level22}Structural morphology}

Superconducting aluminum films were grown in our state-of-the-art cryogenic molecular beam epitaxy system (Scienta Omicron EVO 50) as described in the methods section and in \cite{PhysRevApplied.23.034025}. This system has a base pressure lower than $5\times 10 ^{-11}$ mbar and the ability to cool-down substrates to a base temperature of 4.8 K. The samples grown in this study were deposited at a nominal temperature of 6.2 K to 6.4 K. After deposition, they were transferred to the preparation chamber where they were oxidized while still cold with high purity oxygen (99.994\%). This was done as a way to reduce atom surface mobility and preserve the structure of aluminum that results from cryogenic growth \cite{doi:10.1126/science.aba5211}. Unless otherwise noted, the results presented in the main text of the manuscript correspond to 50 nm thick films for samples grown at room temperature and 60 nm thick films for samples grown at 6 K as extracted from x-ray reflectivity data. We believe that the difference in thickness of these films that were deposited using the same aluminum effusion cell temperature (1100 °C) for the same amount of time originates from the thermal expansion mismatch of aluminum during warm-up from 6 K to 293 K. This is supported by x-ray reflectivity which shows changes in the critical angle. This allowed us to determine the density of the cryogenically grown film that decreases from its regular value of $2.7\ g/cm^3$ at room temperature to $2.4\ g/cm^3$ when grown at 6 K. 





Cryogenic deposition is monitored \textit{in-situ} using reflection high energy electron diffraction (RHEED) at 15 keV (Fig. 1 (b) and (c)). As expected, we see a difference in the RHEED pattern between the samples grown at 293 K in comparison to the ones grown at 6 K. In the former, we observe a streaky RHEED pattern that is different in the aluminum $<0\bar{1}1>$ and $<2\bar{1}\bar{1}>$ directions (supplementary material) which is indicative of epitaxial growth \cite{10.1063/5.0213941} while in the samples grown at 6 K we observe rings in all directions which indicate polycrystallinity. These changes in the structure can also be measured using atomic force microscopy (AFM) which shows a dramatic difference in the surface morphology of both films. In the case of the films grown at 293 K (Fig. 1(d)) we see large domains that are well aligned and are around one hundred nanometers in size. In the case of aluminum grown at 6 K (Fig. 1(e)) we observe the presence of small grains that are tens of nanometers. We also see the presence of thin fissures that extend for hundreds of nanometers across the film. We note that thinner films grown at cryogenic temperatures (5 and 10 nm) that were also oxidized cold have similar surfaces with similar grain sizes but with the absence of these fissures. This is consistent with the absence of dislocations in films that are around the critical thickness (6 nm) of aluminum grown on sapphire \cite{MATTHEWS1974118, 10.1063/1.97637}. In thicker films, the fissures become larger and go deeper into the film (see supplemental material). This makes us believe that the fissures appear as a result of thermal expansion mismatch between sapphire and aluminum occurring during the sample warm-up from 6 K to room temperature.

\begin{figure}
\includegraphics[width=130mm]{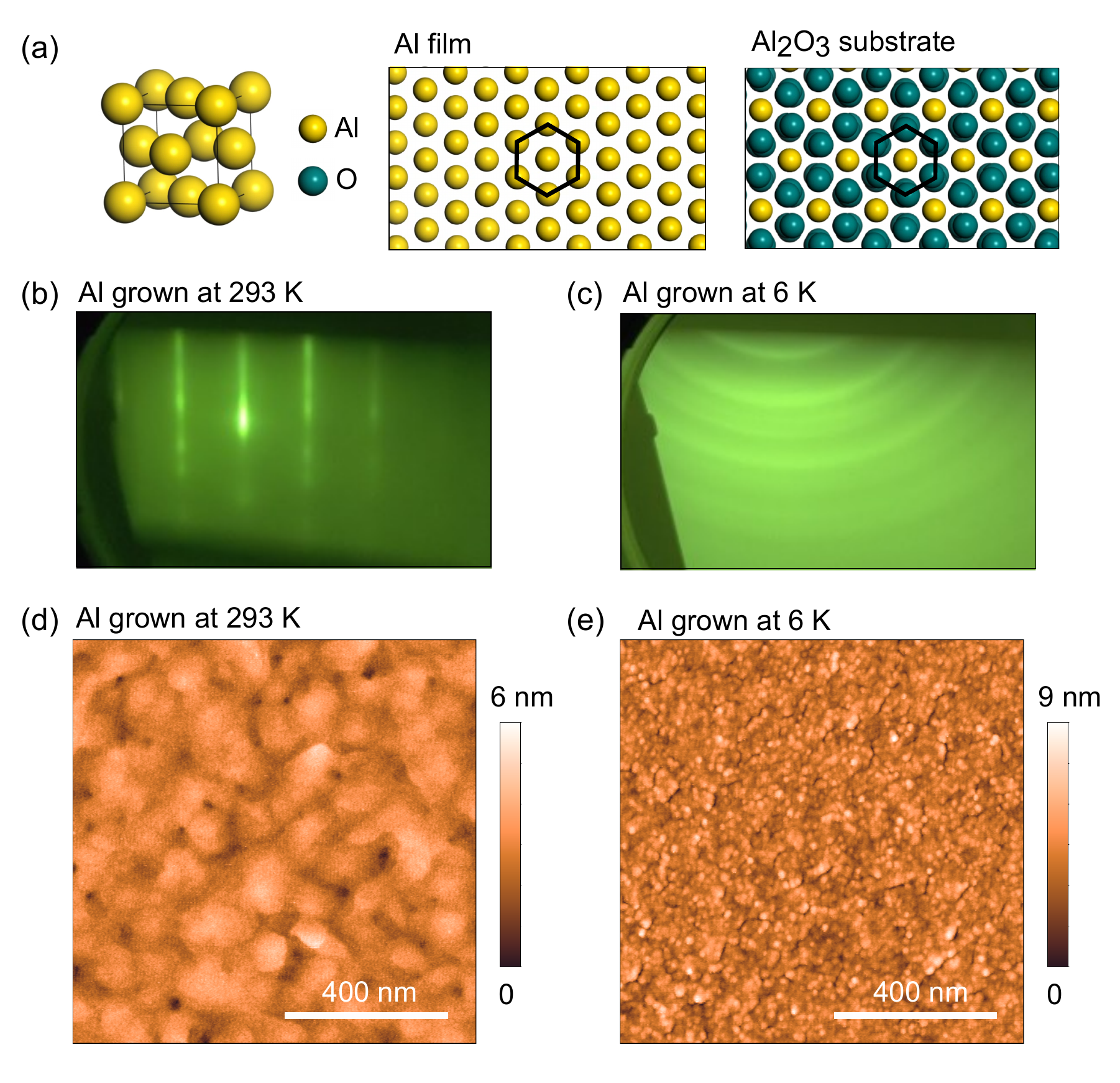}
\caption{\label{fig:1} (a) Crystal structure of aluminum and schematic showing the (111) plane of aluminum and the c-plane of sapphire. Reflection high energy electron diffraction pattern of aluminum grown at (b) 293 K and (c) 6 K. The former corresponds to the Al $<0\bar11>$ crystalline orientation while the latter showed rings in all directions. Atomic force microscopy image of the surface of aluminum grown at (d) 293 K and (e) 6 K.}
\end{figure}



The films were characterized on a macroscopic scale via x-ray diffraction. We performed coupled $\mathrm{2\theta-\omega}$ scans on Al grown at different temperatures and with different thicknesses (Fig. 2 and supplemental material). In all films we observe the expected substrate peaks corresponding to $\mathrm{Al_2O_3}$ (0006) and (00012). In the case of aluminum grown at 293 K (Fig. 2 (b)), we observe the presence of the Al (111) and Al (222) peaks as expected from an epitaxial Al film grown on c-plane sapphire. In the case of aluminum grown at 6 K (Fig. 2 (d)), we see a reduction in the intensity of the Al (111) peak and a complete absence of the Al (222) peak. This is consistent with having some crystalline grains that are tilted from the out-of-plane $<111>$ direction. We have studied the formation of these domains as a function of thickness in cryogenically grown films (see supplementary material) and see the presence of the Al (111) peak at all thicknesses (5, 10 and 60 nm). This indicates that there are $\{ 111 \}$ oriented domains even at very low thicknesses. To understand the microscopic structure of Al, we use high-angle annular dark-field scanning transmission electron microscopy (HAADF-STEM). In aluminum grown at 293 K (Fig. 2 (a)) we see an epitaxial film with well-defined rows of atoms on top of a crystalline $\mathrm{Al_2O_3}$ substrate. In the case of aluminum grown at 6 K (Fig. 2(c)), we see the presence of polycrystalline grains that have a well-defined crystal structure oriented in different directions and are tens of nanometers in size. We see that some of these domains are aligned (with a minor tilt) to the out-of-plane $<111>$ direction. This is consistent with RHEED, atomic force microscopy, and x-ray diffraction characterization. We have also done compositional analysis using energy-dispersive x-ray spectroscopy (supplemental material) and see the expected composition of the film and the substrate. In both films we see the presence of a 1 nm thick amorphous layer at the interface between Al and $\mathrm{Al_2O_3}$. This indicates that greater emphasis should be placed on understanding the growth dynamics during the initial stages of film formation. A first step would be to implement more careful substrate preparation with different cleaning procedures \cite{Place2021} or with higher temperature anneals in a different atmosphere \cite{10.1063/1.114313} as a way to improve this interface.

\begin{figure}
\includegraphics[width=150mm]{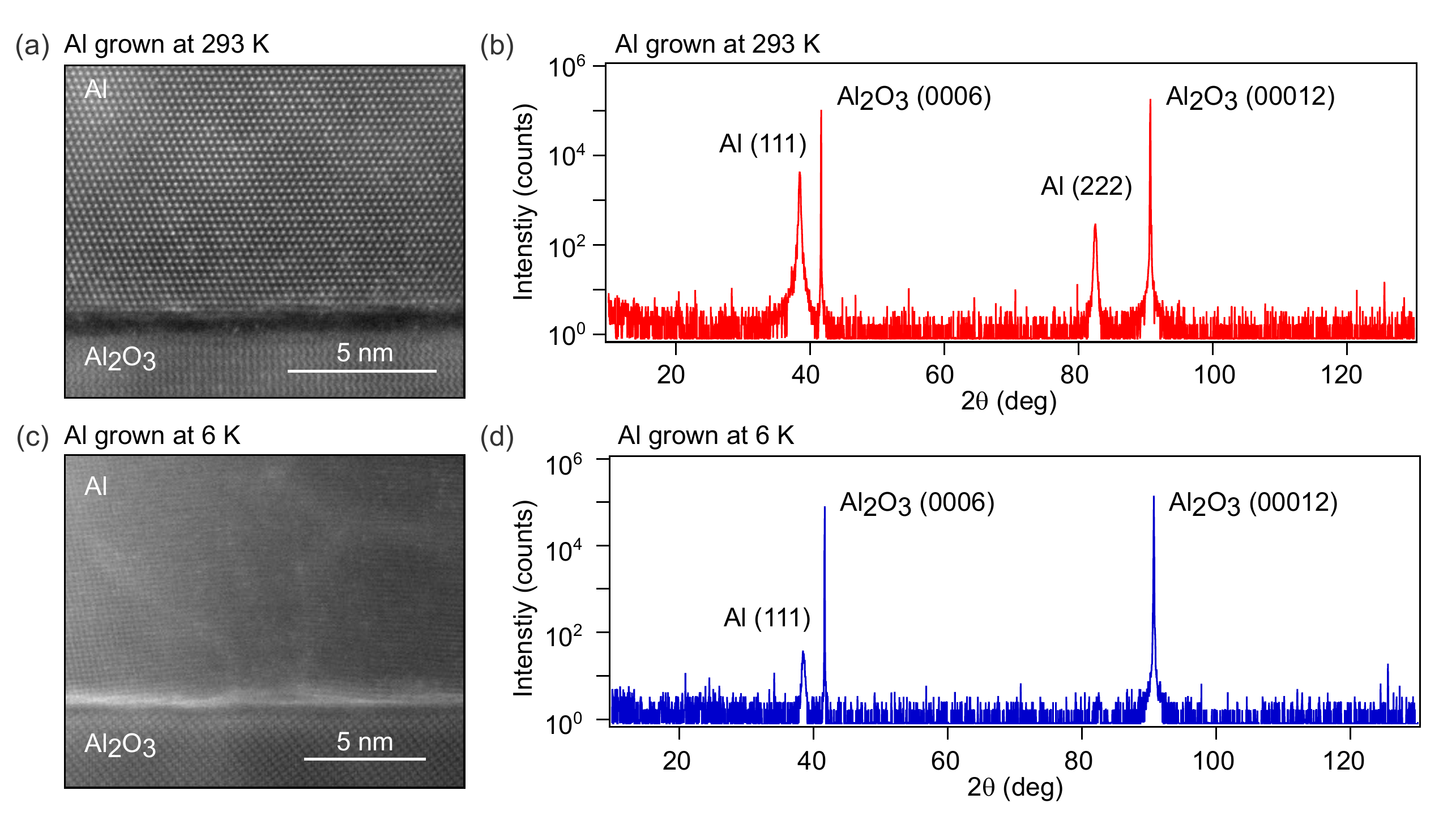}
\caption{\label{fig:2} High angle annular dark field-scanning transmission electron microscopy image of the Al film grown at 293 K (a) and 6 K (c). X-ray diffraction $\mathrm{2\theta - \omega}$ scan of Al grown at 293 K (b) and 6 K (d). }
\end{figure}

\subsection{\label{sec:level23} Optical properties}

The optical properties of aluminum grown at cryogenic temperatures were extracted from spectroscopic ellipsometry measurements. We observed that when grown at cryogenic temperatures, the color of aluminum changes from fully reflective to yellow (Fig. 3(a)). This effect becomes more apparent when looking at the reflectance data of the sample calculated from spectroscopic ellipsometry (Fig. 3(b)). We see that in the cryogenically grown films there is a sharp drop in the reflectance at wavelengths of around 400 nm which corresponds to the absence of the color blue being reflected off the sample. To better understand this phenomenon, we calculated the real and imaginary parts of the pseudo-dielectric function ($<\varepsilon>$) of both films (Fig. 3(c)) and used ellipsometry modeling to understand the physics behind their optical response \cite{fox2001optical,EMARef}. For the case of completely reflective aluminum grown at 293 K, we observe the expected behavior of a metal described by the Drude model with a dielectric function that diverges at low photon energies and goes to zero at photon energies significantly lower than the plasma frequency of aluminum ($\hbar \omega_p=15.2$ eV) \cite{PhysRev.132.1918,fox2001optical}. We also see the presence of a peak at 1.5 eV corresponding to the inter-band transitions of aluminum close to the W and K points in the Brillouin zone \cite{PhysRev.132.1918}. This response can be fitted using the known dielectric function of aluminum. In the case of cryogenically grown yellow aluminum, the behavior is more complex. We first note that there is a decrease in the magnitude of the dielectric function at low incident photon energies. This is due to a lower conductivity in cryogenically grown aluminum due to disorder. This was taken into account in the fit by using a general oscillator model with adjustable broadening to account for increased scattering due to this lower conductivity. We then note that there is the appearance of a second broad feature in the spectra at photon energies between 2.5 eV (496 nm) and 5 eV (248 nm) which causes the change in color. The only way we found to reproduce this response is by modeling the surface of the film using a 5 nm thick effective medium approximation (EMA) layer composed of aluminum with an 8.3 \% void fraction that has a depolarization factor of 0.992 \cite{EMARef}. This reflects the fact that aluminum grown at 6 K develops elongated fissures in the top of the film, as is seen in atomic force microscopy characterization (Fig. 1 (e) and supplemental material). We note that thinner films (5 and 10 nm) that do not have the presence of fissures on their surface have a different pseudo-dielectric function (see supplemental material). Nevertheless, they do not show a feature between 2.5 and 5 eV in their pseudo-dielectric function and do not become yellow. This makes us confident that the presence of fissures on the surface of the film is the reason for the change in color.

\begin{figure}

\includegraphics[width=140mm]{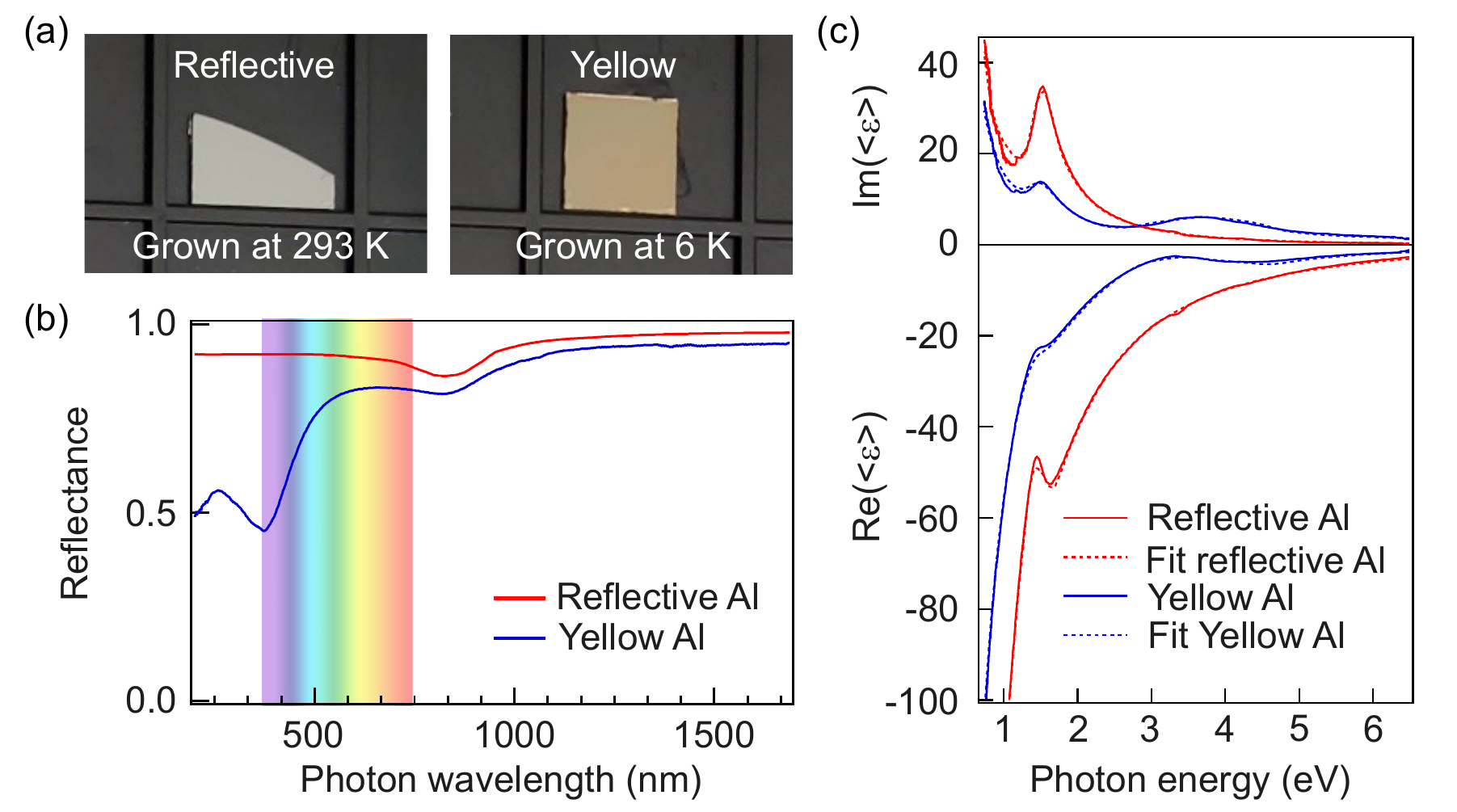}
\caption{\label{fig:3} (a) Pictures of completely reflective aluminum grown at 293 K and yellow aluminum grown at 6 K. (b) Reflectance spectra of reflective and yellow aluminum as a function of incident photon wavelength. (c) Real and imaginary parts of the pseudo-dielectric function of reflective and yellow aluminum. }
\end{figure}


\subsection{\label{sec:level24} Superconductivity and microwave loss}


Electrical transport was measured in aluminum thin films from room temperature to the superconducting regime. Immediately we see a difference in the resistivity of the films as a result of structural differences in them. The resistivity at room temperature of cryogenically grown aluminum ($\mathrm{\rho_{6K}=49.7 \ \mu\Omega\ cm}$)  is significantly higher than that of aluminum grown at room temperature ($\mathrm{\rho_{RT}=10.5\ \mu\Omega\ cm}$). We measured the residual resistivity ratio (RRR) down to 2 K and obtained a value of $RRR_{6K}=1.36$ for cryogenically grown aluminum and $RRR_{RT}=5.64$ for aluminum grown at room temperature. These results are consistent with the cryogenically grown film having smaller grains that produce a higher electron scattering at grain boundaries. We measure the superconducting transition as a function of temperature and magnetic field perpendicular to the sample plane (Fig. 4(a)). We find that samples grown at 6 K have an enhanced critical temperature ($T_C$) and critical field ($H_C$) of $\mathrm{T_C=1.57\ K}$ and $\mathrm{H_C=685\ Oe}$ compared to aluminum grown at 293 K which has $\mathrm{T_C=1.19\ K}$ and $\mathrm{H_C=43\ Oe}$. This increase is consistent with similar results in granular aluminum, in which particle size can enhance superconductivity due to confinement \cite{10.1063/5.0250146Dome,PhysRevLett.100.187001,Bose2014ua,PhysRevB.93.100503}. The values of aluminum grown at room temperature are similar to ones reported in the literature for thin films \cite{PhysRevB.5.3558}. We estimate the superconducting gap and the coherence length of Cooper pairs ($\xi$) under out-of-plane magnetic field in the dirty limit using $\Delta_{SC}=1.76K_BT_C$ and $\xi=\sqrt{\frac{\Phi_0}{2\pi H_c}}$. Where $\Delta_{SC}$ is the superconducting gap, $K_B$ is the Boltzmann constant, and $\Phi_0$ is the magnetic flux quantum. We find that for aluminum grown at 293 K these values are $\mathrm{\Delta_{SC}=181\ \mu eV}$, $\mathrm{\xi=271\ nm}$ and for aluminum grown at 6 K they are $\mathrm{\Delta_{SC}=240\ \mu eV}$, $\mathrm{\xi=69\ nm}$. This indicates that the coherence length of Cooper pairs in aluminum grown at 6 K is similar to the film thickness.


We then investigate the microwave dielectric loss in aluminum grown on sapphire by fabricating capacitive coupled superconducting resonators with a feed line width of $\mathrm{w=5.7\ \mu m}$ and a gap spacing of $\mathrm{s=3.2\ \mu m}$ (Fig. 4(c)). We designed the resonators to operate at a frequency range of 5.5 to 6.5 GHz with a capacitive coupling quality factor of $\mathrm{Q_C=3\times 10^5}$. We computed the loss on the resonators using the diameter correction method \cite{10.1063/1.3692073}. With this we obtain the internal quality factor at different photon numbers as shown in figure 4(b). Representative low photon number transmission data can be found in the supplementary material. We have measured a total of 14 resonators at high power and 8 of them down to the single-photon regime. We observe that the internal quality factor saturates at low power as expected for a system dominated by two-level system loss. We have summarized our results in figures 4(d) and 4(e). We observe that the quality factor is very similar among all the resonators measured with the sample grown at 6 K having marginally higher magnitude. We find that at high photon number ($\mathrm{\bar{n}=1.6\times10^6}$) the internal quality factor of aluminum grown at 6 K can be as high as $\mathrm{Q_i=1.2 \times 10^7}$ with an average number of $\mathrm{Q_i=7.6 \times 10^6}$ while aluminum grown at 293 K has a maximum quality factor of $\mathrm{Q_i=6.2 \times 10^6}$ and an average of $\mathrm{Q_i=4.5 \times 10^6}$. We average the quality factor below a single photon ($\mathrm{\bar{n}<1}$) and obtain similar results with an average quality factor of $\mathrm{Q_{i,LP}=3.2 \times 10^5}$ for the sample grown at 6 K and $\mathrm{Q_{i,LP}=2.9 \times 10^5}$ for the sample grown at 293 K. The differences in these values might be due to the differences in etching and small variation in dimensions during the fabrication of the resonators. We also observe a shift in frequency of the resonators of 0.85 GHz on average which corresponds to a ratio of the kinetic inductance compared to the geometric inductance of the resonator equal to $L_K/L_{geo}=0.36$ for 60 nm thick films. To determine the magnitude of the kinetic inductance of cryogenically grown aluminum, we performed electromagnetic simulations based on the resonance frequency of the microwave resonators and their geometry (see supplemental material). We obtained a value of $L_K=0.79\ pH/\square$ for the kinetic inductance of cryogenically grown aluminum and a sapphire dielectric constant of $\varepsilon_r=11.25$. Since kinetic inductance is proportional to the square resistance ($R_N$) of the film in the normal state ($L_K=\frac{\hbar R_N}{\pi\Delta_{SC}}$), we expect this ratio to be larger in thinner films \cite{10.1063/1.4737408,PhysRevApplied.20.044021,10.1063/5.0017749}.

\begin{figure}
\includegraphics[width=130mm]{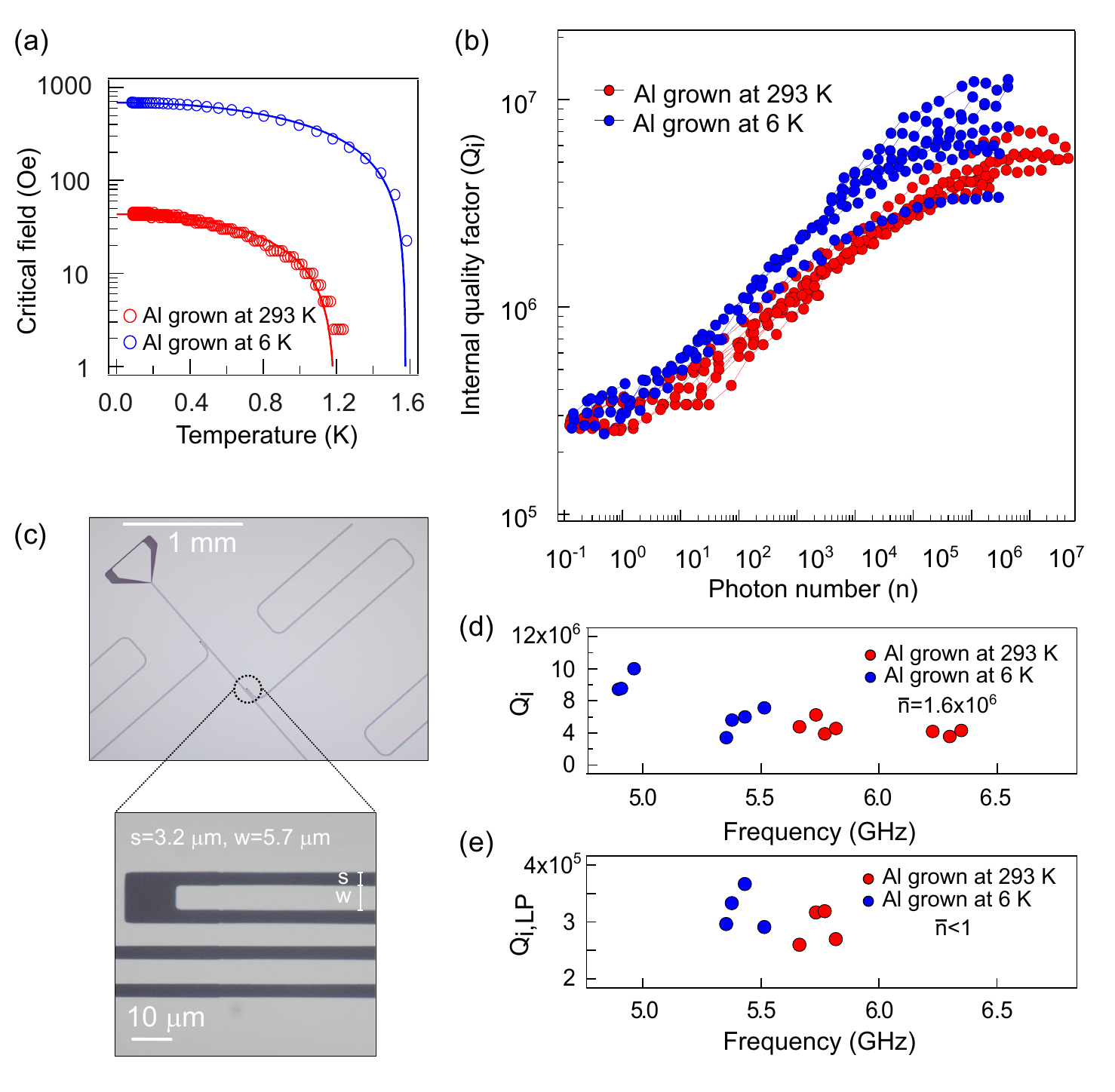}
\caption{\label{fig:5} (a) Critical field as a function of temperature for Al grown at 293 K and at 6 K. (b) internal quality factor as a function of photon number in the cavity for the resonators measured in this work. (c) Optical microscopy images of the microwave resonators measured in this work. The gap width is $\mathrm{3.2\  \mu m}$. Internal quality factor as a function of frequency measured at high photon number $\mathrm{\bar{n}=1.6\times10^6}$ (d) and low photon number $\mathrm{\bar{n}<1}$ (e).}
\end{figure}

The similarity in the quality factor between all resonators indicates that structural disorder, grain size, and the absence of epitaxy in the superconductor have a marginal effect in the $\mathrm{Q_i=3 \times 10^5}$ regime. This is surprising considering that the surface of the film grown at 6 K has fissures that are significant enough to change its optical properties. It indicates that there are other sources of decoherence (like the substrate/metal interface or the sidewalls of the resonator) that might limit the quality factor of the resonators. In thicker films, it has been reported that aluminum grain boundaries can allow more oxide to move to the metal-substrate interface, which had a direct impact on the performance of transmon qubits fabricated on silicon \cite{Biznárova2024}. We expect this phenomenon to be suppressed in this study since the films were grown on sapphire substrates, had lower thicknesses and were oxidized while still cold. Understanding the precise mechanisms of oxidation under cryogenic temperatures and the correlation between optical properties and microwave loss tangent is beyond the scope of this work, but it can provide more strategies on how to improve qubit and resonator performance in future studies.



\section{Conclusion}


We have shown that cryogenic growth dramatically changes the structural, optical, and electrical properties of aluminum deposited on c-plane sapphire. When grown at 6 K aluminum thin films become polycrystalline with grains that are tenths of nanometers in size and have predominantly $\{111\}$ texture. If the film is thicker than 60 nanometers, we start to observe fissures in the surface of the film which change the pseudo-dielectric function of aluminum turning the films yellow. Structural disorder in the form of smaller grain size enhances the superconducting properties of aluminum and increases its critical temperature ($\mathrm{T_C=1.57\ K}$) and critical field ($\mathrm{H_C=685\ Oe}$) without affecting the quality factor ($\mathrm{Q_{i,LP}=3.2\times10^5}$ at low power) of superconducting microwave resonators in a significant way. Moreover, the presence of disorder enhances the kinetic inductance of aluminum ($L_K=0.79\  pH/\square$ in 60 nanometer thick films) which can be advantageous for technological applications like single-photon detectors, parametric amplifiers and quantum bits. The ability to control structural properties via cryogenic deposition in an ultra high vacuum environment with \textit{in-situ} characterization opens the door to study fundamental scientific questions like the interplay between superconductivity and Anderson localization and superconductor thin film growth kinetics in a completely new regime. 

\section{Methods}

\subsection{Cryogenic aluminum deposition}

We describe the synthesis of superconducting aluminum films under cryogenic conditions. To prepare the sapphire substrates for deposition, we etch them using piranha solution $H_2SO_4:H_2O_2$ at a 3:1 concentration for 5 minutes and immediately load them into an ultra-high vacuum chamber where we anneal them at a temperature between 700 °C and 900 °C for 1 hour. We then transfer the samples without breaking vacuum to our state-of-the-art cryogenic molecular beam epitaxy system (Scienta Omicron EVO 50) which has a base pressure lower than $5\times 10 ^{-11}$ mbar. Here we can cool down the substrates to a base temperature of 4.8 K, being measured by a silicon diode located on the substrate manipulator as described in \cite{PhysRevApplied.23.034025}. The cooling power is provided by a custom substrate manipulator connected to a two-stage closed-cycle Gifford-McMahon cryocooler. The samples grown in this study were deposited at a nominal temperature of 6.2 K to 6.4 K, which is higher than the base temperature of our manipulator due to heat coming from the effusion cell impinging on the sample. During evaporation the substrate temperature can increase up to 3 degrees. We evaporate aluminum at a cell temperature between 1100 °C and 1140 °C using a high purity source (6N5) from a cold-lip effusion cell. This corresponds to a growth rate of 0.58 nm/min to 1.3 nm/min. Results for other thicknesses can be found in the supplemental material. After growth we immediately transfer the samples while they are still cold to a load lock chamber where we oxidize the film using high purity oxygen (99.994\%) for 15 minutes at 10 torr. After oxidation, we pump out the oxygen and let the sample naturally warm-up to room temperature for several hours. We have performed \textit{in-situ} x-ray photoemission spectroscopy (see supplemental material) after growth but before the oxidation step and we observe no other elements in the film besides aluminum.  

\subsection{Optical ellipsometry}

Ellipsometry measurements were performed using a variable-angle spectroscopic ellipsometer (J.A. Woollam M2000-DI) at incidence angles of 55°, 65°, and 75°. Unless otherwise specified, the data presented in this work correspond to the 65° measurements. The spectral range for all measurements spanned wavelengths from 193 nm to 1650 nm.

\subsection{Microwave resonator fabrication}

The devices shown in this study have been patterned using standard optical lithography techniques with different etching procedures. For aluminum grown at room temperature we used reactive ion etching with $\mathrm{BCl_3/Cl_2}$ precursors while for aluminum grown at cryogenic temperatures, we used AZ 400k 4:1 developer as an etching agent which is capable of completely removing the aluminum film during resist development.   

\subsection{Direct current electrical transport}

 Direct current (dc) electrical transport was measured in a 4 point geometry using an adiabatic demagnetization refrigerator (base temperature 60 mK). Ti/Au contacts were deposited \textit{ex-situ} using a hard mask to avoid lithography and lift-off. The values of the resistivities mentioned in the main text were obtained from the slope of I-V curves measured at different temperatures. Immediately we see a difference in the resistivity of the films due to structural disorder. We measured the whole resistance, temperature, and magnetic field phase space in the superconducting regime to extract the critical field as a function of temperature (Fig. 4(a)).

\subsection{Duration and volume of study}

The study was spread over 2023-2025 and included a total of 29 aluminum films grown on c-plane sapphire substrates. The structural and optical  characterization included around 110 different sets of measurements. DC electronic transport was measured in two different films and included around 950 scans. We measured 3 different chips with 22 resonators in total for microwave characterization. Among these, 2 chips with 14 resonators are presented in this work. Approximately 1100 microwave data sets were collected.

\section{Acknowledgments}

The cryogenic temperature superconductor synthesis was funded by the U.S Department of Energy under award number DE-SC0025017. The microwave characterization was supported by the New and Emerging Qubit Science and Technology (NEQST) Program initiated by the U.S. Army Research Office (ARO) under Grant No. W911NF2210052.  Ellipsometry characterization was supported by the NMSU/UCSB partnership under award No. NSF PREM DMR-24223992. We acknowledge the support of the University of California, Santa Barbara (UCSB) National Science Foundation (NSF) Quantum Foundry through Q-AMASE-i Program via Award No. DMR-1906325. This work also utilized shared facilities provided by the UCSB MRSEC (NSF DMR–2308708) and the Nanotech UCSB Nanofabrication facility. Certain commercial equipment, instruments, or materials are identified in this paper to foster understanding. Such identification does not imply recommendation or endorsement by the National Institute of Standards and Technology, nor does it imply that the materials or equipment identified are necessarily the best available for the purpose.





\providecommand{\noopsort}[1]{}\providecommand{\singleletter}[1]{#1}%

 \end{document}